\documentclass[%
preprint,
 amsmath,amssymb,
prl,
]{revtex4-1}

\usepackage{graphicx}
\usepackage{dcolumn}
\usepackage{bm}
\usepackage{amsmath}
\usepackage{amssymb}
\usepackage{mathrsfs}
\usepackage{epsfig}

\def\comment#1{}

\begin{document}

\preprint{APS/123-QED}

\title{Excitation of high frequency voices from intermediate-mass-ratio inspirals with large eccentricity}
\thanks{This work is support by NSFC No. U1431120 and No.11273045  }%

\author{Wen-Biao Han$^{1,2}$, Zhoujian Cao $^{3}$, Yi-Ming Hu $^4$}
\email{ wbhan@shao.ac.cn}
\affiliation{%
 1. Shanghai Astronomical Observatory, Shanghai, 200030, China   \\
 2. School of Astronomy and Space Science, University of Chinese Academy of Sciences,  Beijing 100049, China \\
 3. Beijing Normal University, Beijing, China, 100875 \\
 4. TianQin Research Center for Gravitational Physics, Sun Yat-Sen  University, Zhuhai, Guangdong, 519082, China}

\date{\today}

\begin{abstract}
The eccentricity can be still large in the final stage of large-mass-ratio-inspiral event.  Modified gravity theories generically predict a violation of Lorentz invariance, which may lead to a dispersion phenomenon for propagation of gravitational waves. In this Letter, we demonstrate that this dispersion will induce an observable deviation of waveforms, if the orbital eccentricity is considerable. The mechanism is that a lot of waveform modes with different frequencies will be emitted at the same time due to the existence of eccentricity. During the propagation, because of the dispersion, the arrive time of different modes will be different, then produce the deviation and dephase of waveforms comparing with general relativity.  The dispersion phenomena revealed in this Letter, can be observed by LISA, Taiji, Tianqin and even advanced LIGO and Virgo.   
\begin{description}
\item[PACS numbers] 04.70.Bw, 04.80.Nn, 95.10.Fh
\end{description}
\end{abstract}

\pacs{04.70.Bw, 04.80.Nn, 95.10.Fh} 
\maketitle

\section{Introduction}

The first detection of gravitational wave (GW), GW150914, was made in September 2015 with the Advanced LIGO (aLIGO) detectors \cite{ligo16a}.
Subsequently, more signals GW151226 and GW170104, together with a likely candidate LVT151012, were detected \cite{ligo16c,ligo17b}.
All such systems are believed to be originated from the merger of binary black hole systems, with no black hole being more massive than $100 M_\odot$. The waveforms from these systems can be accurately modeled by effective-one-body (EOB) theory. The EOB scheme can be understood as a post-Newtonian approach calibrated with black hole perturbation theory and numerical relativity simulations \cite{Buonanno1,Buonanno2,Panyi11,Taracchini12,Taracchini14,purrer14}. 

The LIGO and Virgo collaborations published results from searching for binaries up to hundreds of solar masses \cite{ligo14a,ligo14b}, and recently released their search results on the intermediate mass black hole binaries with total mass less than 1000 solar masses \cite{ligo17}. In principle, the later could be detected by Advanced LIGO and Advanced Virgo (here after aLIGO and AdV) \cite{veitch15}. \comment{However, searching for GWs from a compact object with stellar mass orbiting an intermediate massive black hole with more than one thousand of solar masses may be difficult now because the amplitudes and frequencies of GWs radiated from such kind of systems usually do not reach the sensitive band of the latest updated Advanced LIGO and Advanced Virgo (here after aLIGO and AdV) over a range of frequencies from about 10 Hz to several kHz \cite{ligo}.} However, the coalescence of a stellar mass compact object together with an intermediate massive black hole $\gtrsim 1,000 M_\odot$ has not been searched up to now.
Such kind of systems, which are also known as intermediate-mass-ratio inspirals or IMRIs, usually emit gravitational waves with frequency lower than the sensitive band of advanced LIGO, which typically ranges from about 10 Hz to several kHz \cite{ligo}. The frequencies of these GWs during the final stage ($\sim$ $\mathcal{O}(1)$ Hz) are also out of the sensitive band of the space based detectors like LISA \cite{elisa}, Taiji \cite{taiji} and TianQin \cite{tianqin} etc, which typically span from 0.1 mHz to 0.1 Hz. 

The typical masses of the massive black holes in IMRIs are usually around $\mathcal{O}(10^2) \sim \mathcal{O}(10^4)$ solar masses \cite{davidbook}. These black holes are believed to exist in low-luminosity active galactic nuclei,  globular clusters \cite{merritt13} and some ultra-luminous X ray sources \cite{maccarone07}.  In the present paper, we are interested in the intermediate massive black holes with more than 1000 $M_\odot$. \comment{Central Black holes with several thousands of solar masses will produce too low frequency GWs for aLIGO} Central black holes with several thousands of solar masses would have frequencies too low to be observed by the current ground-based GW detectors. However, the binaries with total masses $O(10^2) M_\odot$ can be observed by aLIGO in circular orbit cases \cite{ligo}. Therefore, in this work we investigate on the range in between, where the mass-ratio is assumed as $m_2/m_1 \sim 10^{-3}$, the mass of the central black hole is around 1000-2000 $M_\odot$, and the small compact body is assumed as white dwarf, neutron star or stellar black hole. In the present paper,  we focus on the detectability of such IMRIs for aLIGO and AdV, and thus leave aside the event rates for future investigation, however interested readers are encouraged to check relevant details in \cite{gair11}. 

This paper is organized as follows. In the next section, we introduce the EOB formalisms for the two-body dynamics. In the section III, we study the IMRIs with large eccentricities and their dominate GW frequencies. Possibility of observing the GWs from IMRIs using ground-based detectors is discussed in section IV. Finally, we give remarks and conclusions.  

\section{Effective-One-Body formalism for eccentric orbits}
The EOBNR model \cite{Taracchini14} has been adopted successfully in the GW signal search of LIGO \cite{ligo16c}. In addition to the test particle approximation, the EOB dynamics includes the mass-ratio correction and spin.  The mass-ratios  of the IMRIs studied in this paper are around $10^{-3}$. In order to include the mass-ratio correction, we employ the EOB formalism to calculate the orbital motion. Many researches have proved the effectiveness of the EOB dynamics for gravitational two-body systems (see e.g. \cite{damour12} ). In the present paper, we use the same EOB dynamical formalisms of the EOBNR model \cite{Taracchini14}. 

\comment{Note that $q$ is the so-called deformed-Kerr spin parameter which is
not exactly equal to the spin parameter of Kerr black hole itself, because $q\equiv |\boldsymbol{S}_\text{Kerr}|/M^2\neq|\boldsymbol{S}_1|/m_1^2$.The values of $\omega^{fd}_1$ and $\omega^{fd}_2$ given by a
preliminary comparison of EOB model with numerical relativity
results are about -10 and 20 respectively \cite{Rezzolla08,Barausse09}.}

The EOB Hamiltonian takes the form \cite{Buonanno1,Buonanno2}
\begin{align}
H_{\text{EOB}}=M\sqrt{1+2\nu(H_{\text{eff}}/\mu-1)}.
\end{align}
Here we define the total mass $M=m_1+m_2$, reduced mass $\mu=m_1m_2/M$ and symmetric mass ratio $\nu=\mu/M$, where $m_1$ and $m_2$ are the masses of the two objects of the binary (we always assume $m_1 > m_2$).The central black hole has spin $S_1 = q M^2$, where $q$ is the effective dimensionless spin parameter of the Kerr black hole. For the smaller compact object, the spin magnitude $S_2 \lesssim \mu^2/\mu M \ll 1$ \cite {hartl03}. Such a small spin will not produce considerable effects in a short time-scale analysis. For simplicity, we omit the spin of the small object: $S_2=0$. However, The effective Hamiltonian $H_{\text{eff}}$ is no longer the Hamiltonian of a non-spinning (NS) test particle $H_\text{NS}$, because even if $S_2=0$ the effective spin is not zero \cite{Barausse10,Barausse11}. Then the effective Hamiltonian should describe a spinning test particle in the deformed Kerr metric \cite{Taracchini12}
\begin{align}
H_{\text{eff}}=H_\text{NS}+H_\text{S}-\frac{\mu}{2Mr^3}S_*^2, \label{effectivehamilton}
\end{align}
where the first part is just the Hamiltonian of a non-spinning particle in the deformed-Kerr metric, and $S_*$ is the effective spin of the particle. The second term includes the spin-orbit and spin-spin couplings. All these quantities involved in the above equations can be found in  \cite{Nagar07,Damour07,Damour09,Bernuzzi10, Bernuzzi11a, Bernuzzi11b, Barausse09} and references inside. 

The conservative dynamical equations in the equatorial plane are
\begin{align}
\frac{dr}{dt}&=\frac{\partial H_\text{EOB}}{\partial {p_r}}\,,\quad \frac{d\phi}{dt}=\frac{\partial H_\text{EOB}}{\partial p_\phi} \,,\label{rdot}\\
\frac{d p_r}{dt}&=-\frac{\partial H_\text{EOB}}{\partial r}\,,\quad \frac{dp_\phi}{dt}=0 \,.\label{pdot}\end{align}

\comment{
\begin{align}
\dot{r}&=\frac{\partial H_\text{EOB}}{\partial p_r}\,,\label{rdot}\\
\dot{\phi}&=\frac{\partial H_\text{EOB}}{\partial p_\phi} \,,\label{fdot}\\
\dot{p_r}&=-\frac{\partial H_\text{EOB}}{\partial r}+\mathcal{F}_r \,,\label{prdot}\\
\dot{p_\phi}&=\mathcal{F}_\phi \,. \label{pfdot}
\end{align}
}
Where $p_\phi$ is the angular momentum. It is a conserved quantity due to the axis-symmetry of the deformed Kerr metric. When the gravitational waves are considered, radiation reaction $\mathcal{F}_r, \mathcal{F}_\phi$ need to appear in Eqs. (\ref{pdot}). 

In order to describe such an equatorial-eccentric motion, we use the geometric parameters, semi-latus rectum $p$ and eccentricity $e$, to determine the orbital configuration. Similar to \cite{han14}, the procedure of transferring geometric parameters to the initial data for the equation of motion is listed as follows. 

1. with values of $p,~e$ for an orbit configuration, derive periastron $r_p$ and apastron $r_a$ by using the definition $r_a=p/(1-e),~r_p=p/(1+e)$;

2. taking $r_a,~r_p$ into Eq. (\ref{effectivehamilton}) respectively, solve $p_\phi$ and $H_\text{eff}$. Considering the complication of $H_\text{S}$, firstly we solve an approximation value of $p_\phi$ with only $H_\text{NS}$, then obtain the accurate solution by Aitken's iterative method. 

3. with the values of $H_\text{eff}$ and $p_\phi$ at hand, and a set of initial data $r(t=0)=r_p,~\phi(t=0)=0,~p_r(t=0)=0$, numerically integrate the EOB dynamical Eqs. (\ref{rdot}-\ref{pdot}) to obtain the accurate orbits.
\comment{where $\dot{E}$ is the energy flux of gravitational radiation. We use these equations to evolve the orbit instead of the energy balance method used in the previous researches \cite{Glampedakis1,Hughes3,Hughes4,Fujita3}. Attend that here the EOB formalism uses Boyer-Lindquist coordinate \cite{Barausse10}, which is the same with the Teukolsky equation. In this sense, the source term we get from the EOB dynamics can be used in the Teukolsky equation directly.}

\section{IMRIs with large eccentricities}

Within the framework of the EOB formalism, we can calculate the orbital frequency of an equatorial-circular orbit: $\Omega_\phi = \partial H_{\rm EOB} /\partial p_\phi$. 
The dominant GW mode is the (2, 2) mode where $l=m=2$ ($h_{22}$), where $l, ~m$ are harmonic numbers. The corresponding GW frequency is twice the orbital frequency. In Table \ref{wcircular}, we list the frequencies of $h_{22}$ when the small body moves along the innermost stable circular orbit (ISCO) of the system with total mass 1500 $M_\odot$ (solar mass). Here we assume an IMRI system composed of a central Kerr black hole $m_1$ and an inspiralling small object $m_2$ which is restricted on the equatorial plane of central black hole. 
Notice that all listed $h_{22}$ frequencies are below 10 Hz, meanwhile sensitivity of ground-based GW detectors like aLIGO and AdV drop dramatically below 10 Hz (see Figs. \ref{ligo1}, \ref{ligo2}). In other words, such kind of IMRIs in circular orbits are very difficult to observe with current ground-based GW detectors.

\begin{table}[h!]
\caption{The frequency of quadrupole GW $h_{22}$ when the small body on the ISCO. The symmetric mass-ratio is $10^{-3}$, and the total mass $M = 1500 M_\odot$.}
\label{wcircular}
\begin{center}
\begin{tabular}{c| c c c c c c}
\hline \hline
$q$ &$0$ &$0.1$ & $0.3$ & $0.5$ & $0.7$ & $0.9$ \\
\hline
$f_{22}$ (Hz) &$2.93$ &$3.17$ & $3.78$ & $4.68$ & $6.21$ & $9.81$ \\
\hline \hline
\end{tabular}
\end{center}
\end{table}

However, for eccentric orbits, the frequency of strongest GW mode is $\omega_{22\tilde{k}} = 2\Omega_\phi+\tilde{k} \Omega_r$, instead of twice the orbital frequency. Where $\Omega_r \equiv 2\pi/T_r$ and $\Omega_\phi \equiv \Delta \phi/T_r$ are the frequencies of radial and azimuthal motions respectively. $T_r$ is the period of radial motion, $\Delta \phi$ is the swept azimuthal angle after the certain time interval $T_r$ has elapsed. Unlike the circular case, one can not directly obtain $\Omega_r, ~\Omega_\phi$ from Eqs. (\ref{rdot},\ref{pdot}). Alternatively, we perform orbital evolution without radiation reaction, then we determine the period $T_r$ and two orbital frequencies $\Omega_r, ~\Omega_\phi$. Furthermore, these two frequencies are used in the frequency-domain codes of the Teukolsky equation.

We define $\tilde{k}$ as when $k=\tilde{k}$ the $k$-mode energy flux is maximum, while $k$ counts the harmonics created by the linear composition of the two orbital frequencies. When eccentricity $e = 0$,  $\tilde{k}$ equals to 0 exactly (and also the $k$-modes disappear). Usually when $e > 0.1$, $\tilde{k}$ is larger than 0. For example, $\tilde{k} \approx 10$ and $18$ for $e =0.7$ and 0.8 respectively.  In this case, the frequency of dominant GWs is several times larger than the circular orbit cases. We call this phenomenon as the excitation of high frequency GWs due to the large eccentricity. This has been discussed in \cite{peters63} for comparable mass-ratio binaries by post-Newtonian method, and in \cite{glampedakis02, Hughes4, Fujita3, han14} for extreme-mass-ratio inspirals by Teukosky equation.  Ref. \cite{peters63} presented a formula accurately predicts the $k$ values of the dominant energy modes for $l = 2$. This formula also works for extreme-mass-ratio cases, and an approximation formula was given by \cite{Hughes4},
\begin{align}
\tilde{k} \approx \exp{[\frac{1}{2}-\frac{3}{2}\ln{(1-e)}]}\,.
\end{align}
With this expression, we can quantitatively estimate the frequency of dominant mode (hereafter we call it as ``voice") of GWs for different eccentricities.  In Table \ref{wgw}, we find that the frequency $\omega_{22\tilde k}$ grows when the eccentricity becomes larger, and then the signals enter the sensitive band of aLIGO and AdV. Considering that the frequency is inversely proportional to the total mass, one can easily obtain the frequencies for other values of $M$. Though the strength of $h_{22\tilde k}$ decreases as $e$ increases,  the signal still becomes detectable because the sensitivity curves of LIGO detectors drop steeply after the frequency being greater than 10 Hz. This will be demonstrated in the next section.

\begin{table}[h!]
\caption{The frequency of $h_{22\tilde k}$ of the system with symmetric mass-ratio equals $10^{-3}$ and the total mass $M = 1500 M_\odot$. $q$ is the dimensionless spin parameter of the Kerr black hole and $p$ the semi-latus rectum. $p \equiv 2r_{\rm max} r_{\rm min}/(r_{\rm max} +r_{\rm min}$) and eccentricities $e \equiv (r_{\rm max}- r_{\rm min})/(r_{\rm max} +r_{\rm min})$, where $r_{\rm min/max}$ is the the radii of the small body at periapsis/apoapsis. ``$--$'' means unbounded orbits. The unit of numbers listed below is Hz. }
\label{wgw}
\begin{center}
\begin{tabular}{c| c c c c c c c}
\hline \hline
$e$ &$0.1$ & $0.3$ & $0.5$ & $0.7$ & $0.8$  & $0.9$ \\
$\tilde k$ & $0$ & $2$ & $4$ & $10$ & $18$ & $52$  \\
$q=0.90, p=3.10$  &$6.75$ &$8.84$ & $9.93$ & $11.82$  & $--$ & $--$ \\
$q=0.90, p=3.30$  &$6.22$ &$8.30$ & $9.32$ & $11.08$ & $11.86$ & $13.62$ \\

$q=0.95, p=2.79$  &$7.64$ &$10.04$ & $11.30$ & $13.49$ & $14.56$ & $16.85$ \\
$q=0.99, p=2.11$  &$10.59$ &$13.42$ & $15.11$ & $18.11$ & $19.80$ & $23.60$ \\
\hline \hline
\end{tabular}
\end{center}
\end{table}

\begin{figure}
\begin{center}
\includegraphics[height=2.0in]{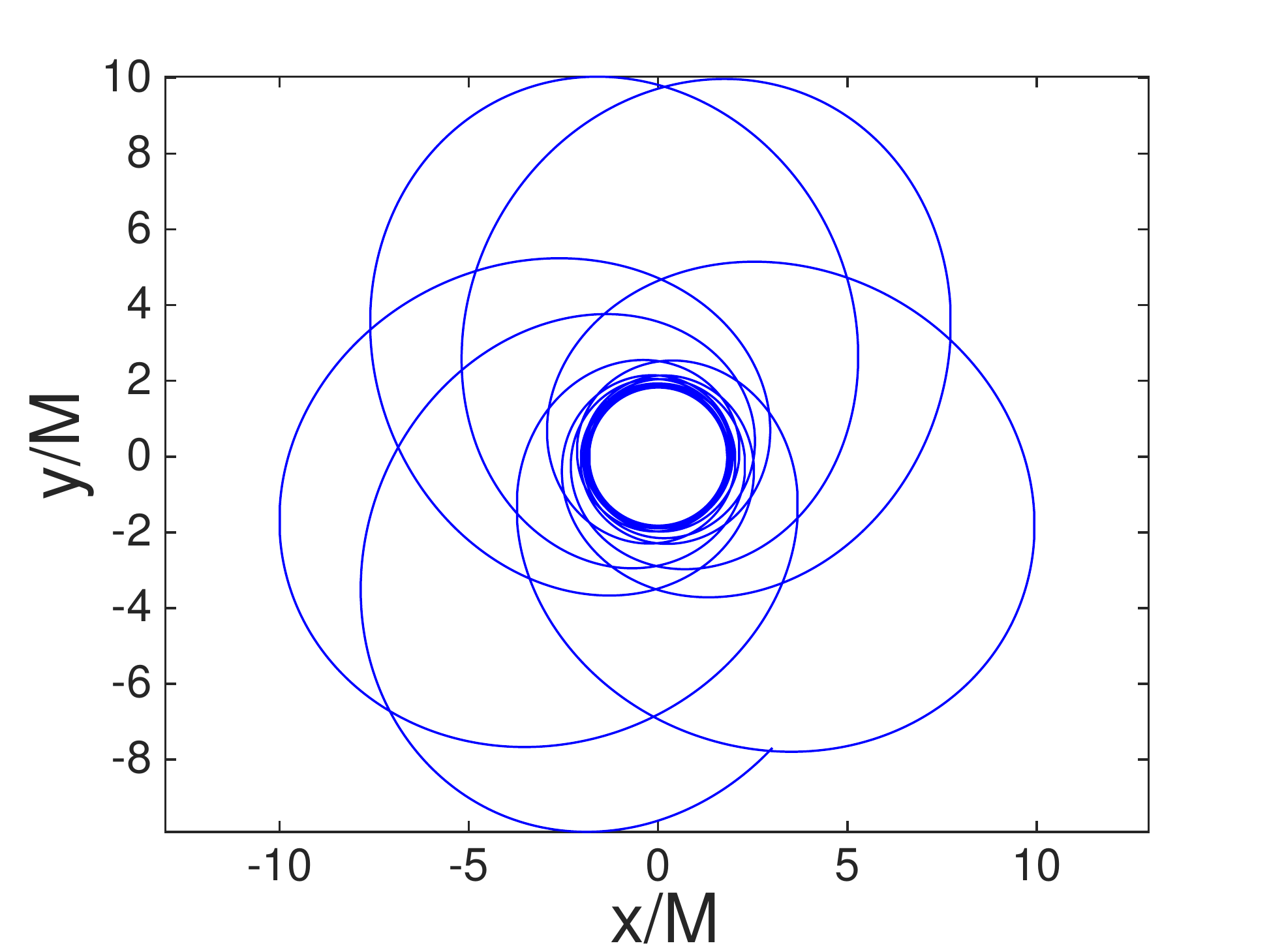}
\includegraphics[height=2.0in]{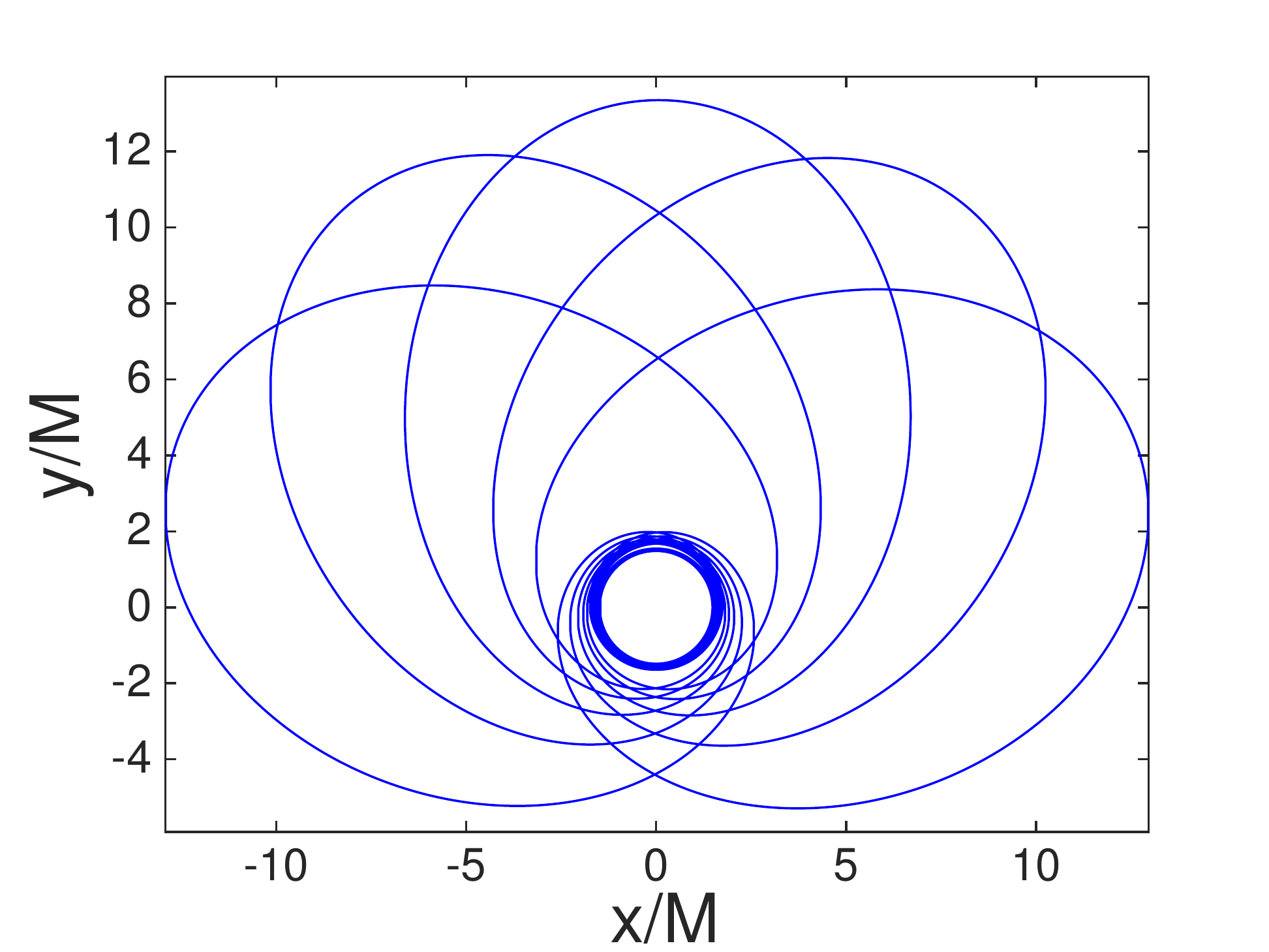}
\includegraphics[height=2.0in]{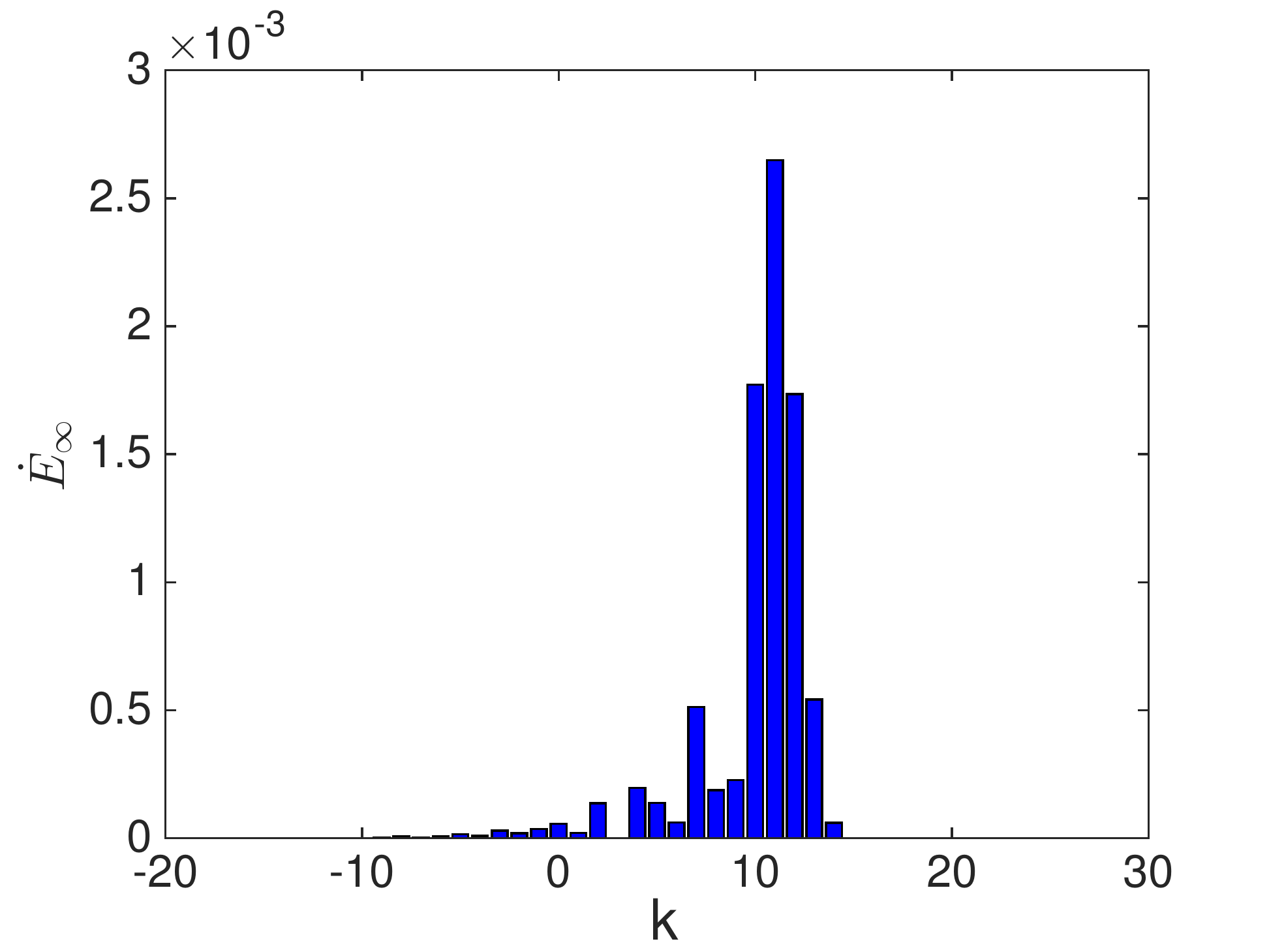}
\includegraphics[height=2.0in]{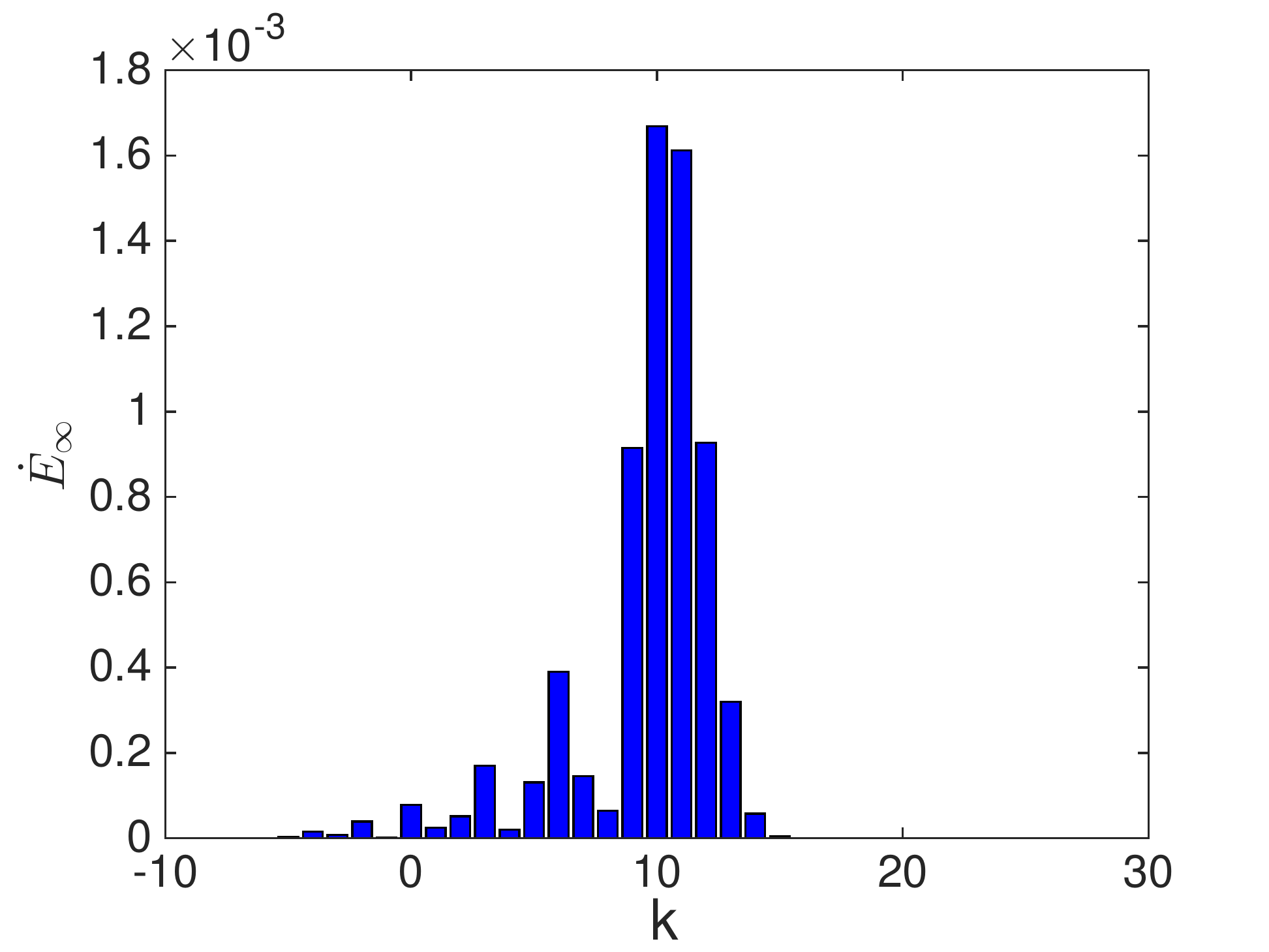}
\caption{Two highly relativistic and eccentric IMRI systems. Top panels show the trajectories of two zoom-whirl orbits. Top-left: $q=0.9$, $p=3.1 M$, $e=0.7$ and $\nu = 10^{-3}$. Top-right: $q=0.95$, $p=2.67 M$, $e=0.8$ and $\nu = 10^{-3}$. Bottom panels show the spectrum of $k$-modes of orbits $a=0.9$, $p=3.1 M$, $e=0.7$ ($\nu = 0$ and $10^{-3}$ in left and right panels respectively).  }  \label{orbitxy}
\end{center}
\end{figure}

In Figure \ref{orbitxy}, two highly relativistic and eccentric orbits are displayed. Features like ``zoom" and ``whirl" could be found in the figure. In the ``zoom-in" phase, the smaller object slowly move towards the center from the apastron,  the ``whirl" phase happens nearby the periastron, where the small object quickly rotates a number of quasi-circular orbits close to the innermost stable bound orbit like a basketball rotates along the basket, afterwards it ``zoom-out" towards the apastron again \cite{glampedakis02}.

In Figure \ref{orbitxy},  we also demonstrate the spectrum of $k$-modes energy flux to infinity ($\dot{E}_\infty$) of two orbits both having the same geometric parameters (semi-latus rectum, eccentricity) $p=3.1 ~M, ~e=0.7$ and the dimensionless Kerr parameter $q=0.9$, but with different mass-ratios $\nu$ equal to 0 and $10^{-3}$ respectively.  The distribution of $k$-modes is very sensitive with respect to the mass-ratio, though the positions of maximum modes are approximately same, i.e., $\tilde{k} = 10$ and 11 respectively.  

\section{The Teukolsky-based gravitational waveforms}
The EOB-Teukolsky (ET) codes employ the EOB dynamical equations to drive the orbits and feed the Teukolsky equation with orbital parameters, then calculate the latter to generate GWs \cite{Han11}.  The Teukolsky equation solver in our ET codes produces gravitational waveforms and energy fluxes, at the same time, the Teukolsky-based energy fluxes can source the EOB dynamics to drive the orbital evolution. For a detailed introduction of numerical methods for the Teukolsky equation, one can see \cite{Teukolsky1,Teukolsky2,Hughes1,Hughes2,Hughes3,Warburton12,Barausse12,Yanbei12,Barton08,Sasaki,Fujita1,Fujita2,MST,Sasaki2} and references inside. For detailed methods of our ET codes for eccentric cases, please see a previous work by one of us \cite{han14}.

Based on the frequency-domain codes, we can accurately calculate the gravitational waves from the eccentric orbits. In principle, these GWs are combined by many individual harmonics (or ``voices'') associated with $l,m$ and $k$ modes. A Fourier decomposition of the GWs can separate these voices (this has be done in the frequency-domain method).  For the circular orbits, the frequency of  $h_{22}$ is fully determined by the orbital frequency $\Omega_\phi$. However, as discussed in the previous section, the situation becomes complicated in the eccentric cases. Because of the participation of $k$-modes, the frequency of the GW mode is a combination of orbital frequencies $\Omega_\phi$ and $\Omega_r$, $\omega_{22k} = 2\Omega_\phi+k\Omega_r$. The maximum mode is located at $k=\tilde{k}$, and $\tilde{k}$ mainly decided by eccentricity.  A wild eccentricity will induce a very large value of $\tilde{k}$, this mechanism shifts the frequency of dominant GW voices higher in comparison with the circular cases. In other words, the eccentricity excites the high mode voice.  

\begin{figure}
\begin{center}
\includegraphics[height=2.0in]{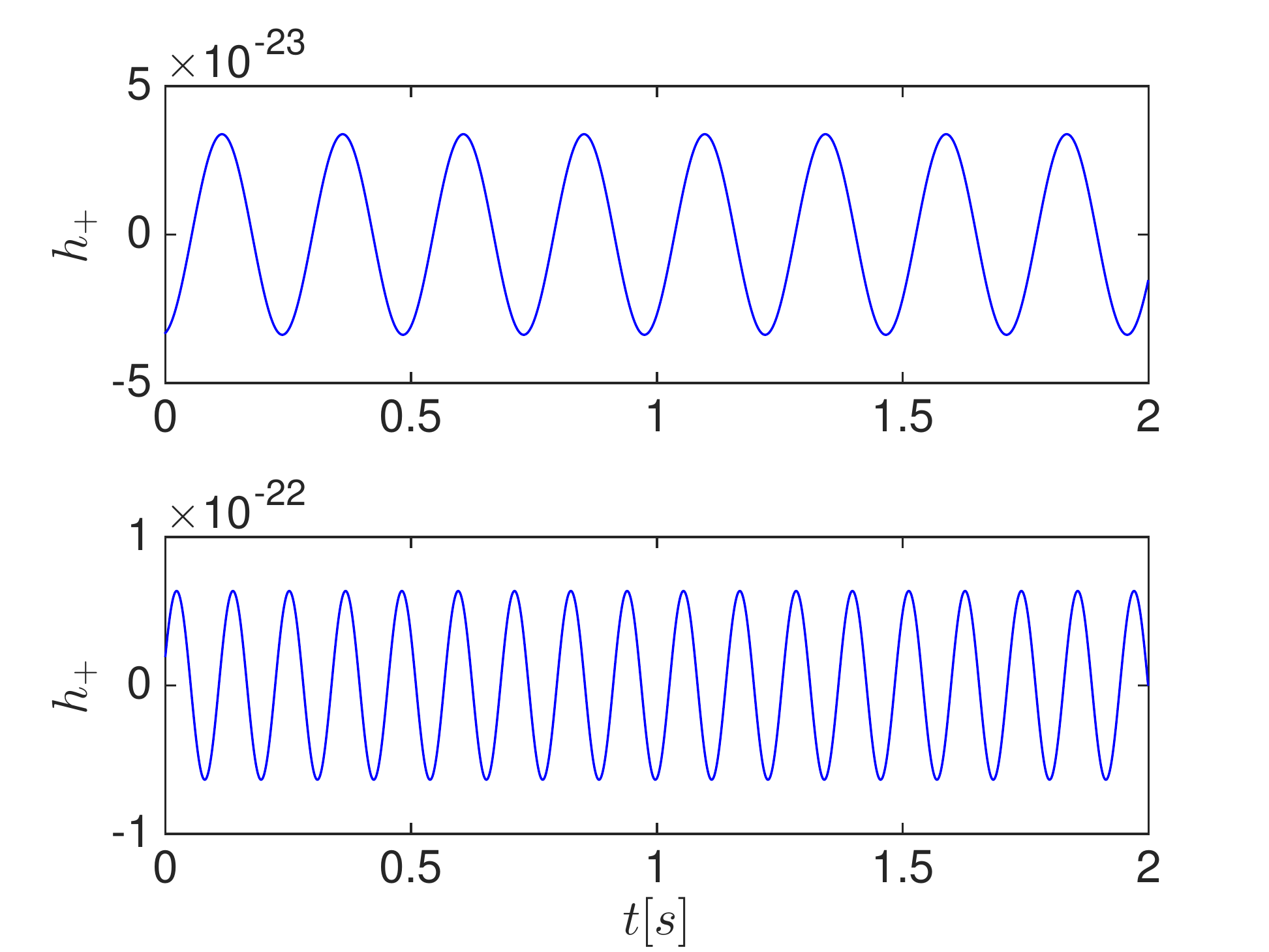}
\includegraphics[height=2.0in]{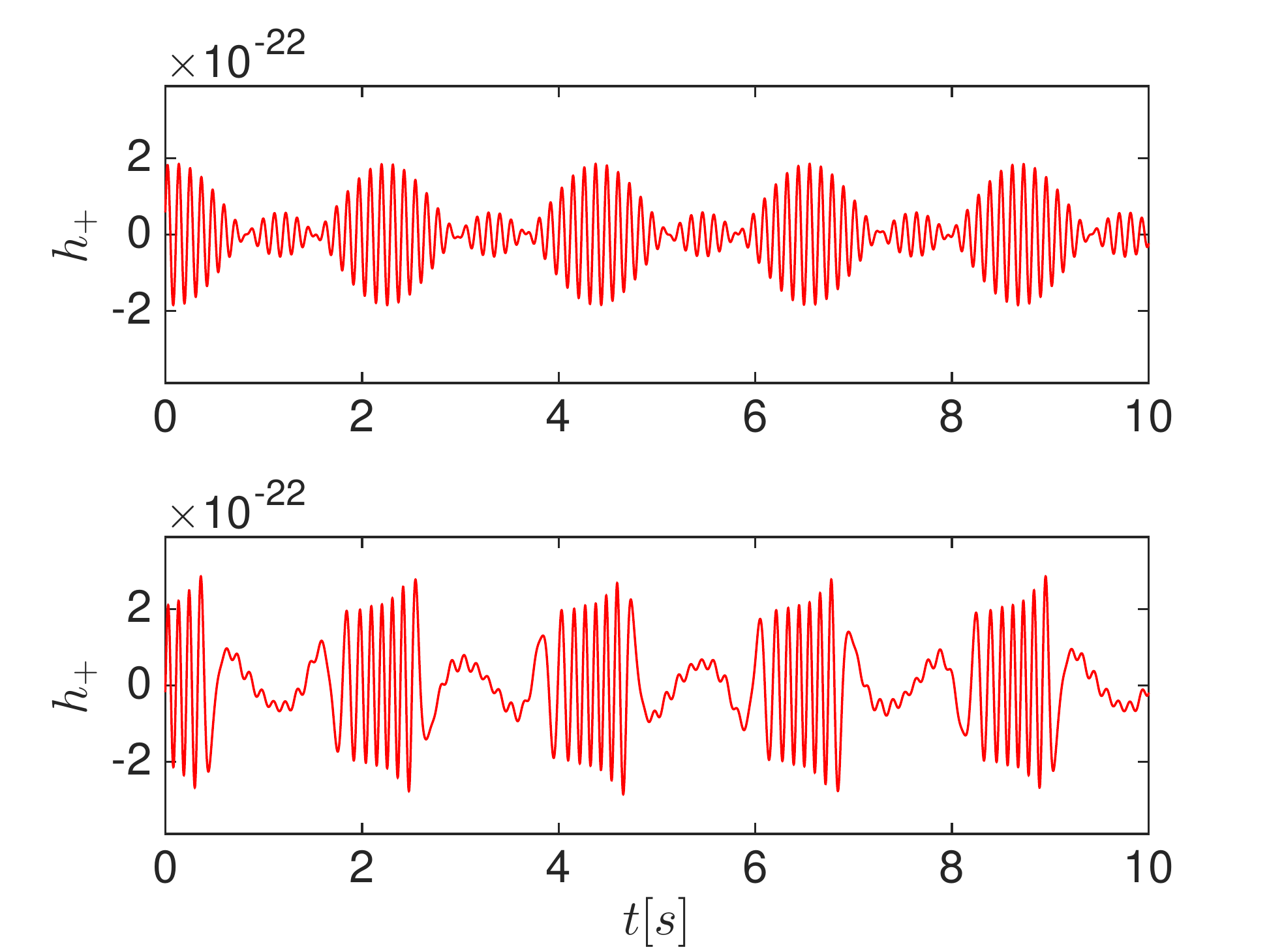}
\caption{Left: The individual harmonics or ``voices" of waveforms associated with $l=m=2$, $k=0$ (top) and $k=\tilde{k} =10$ (bottom); right: the combinations of the $k$ from 9 to 11 (top) and -8 to 20 (bottom) voices, the latter one gives the quadrupole waveform.The orbital parameters are $p=3.1$ and $e=0.7$, the dimensionless spin parameter $q=0.9$, and the mass-ratio is $10^{-3}$. For plotting, the distance is normalized to 100 Mpc, and the mass of the small object is set to 2 $M_\odot$.}  \label{voice1}
\end{center}
\end{figure}

\begin{figure}
\begin{center}
\includegraphics[height=2.0in]{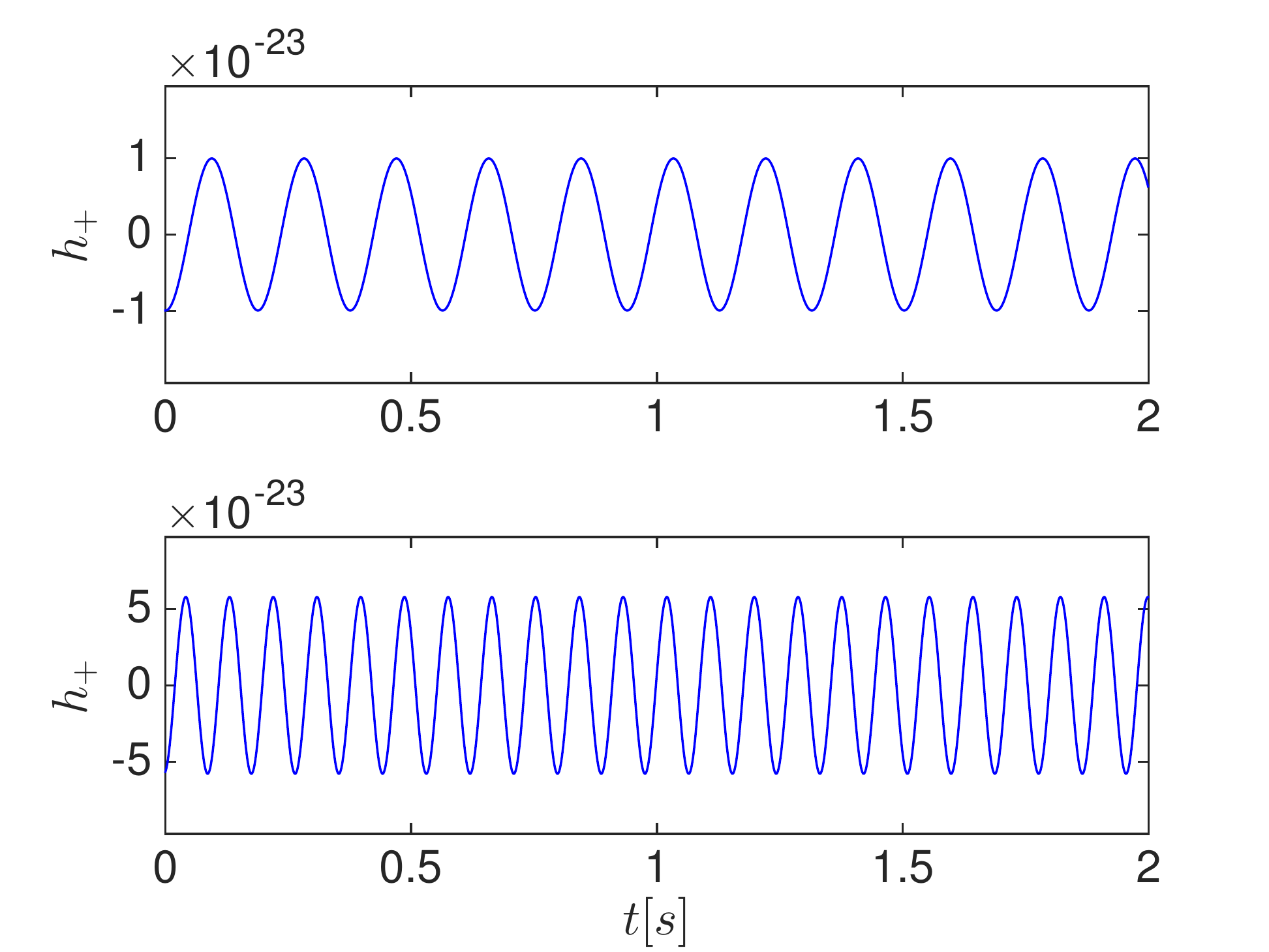}
\includegraphics[height=2.0in]{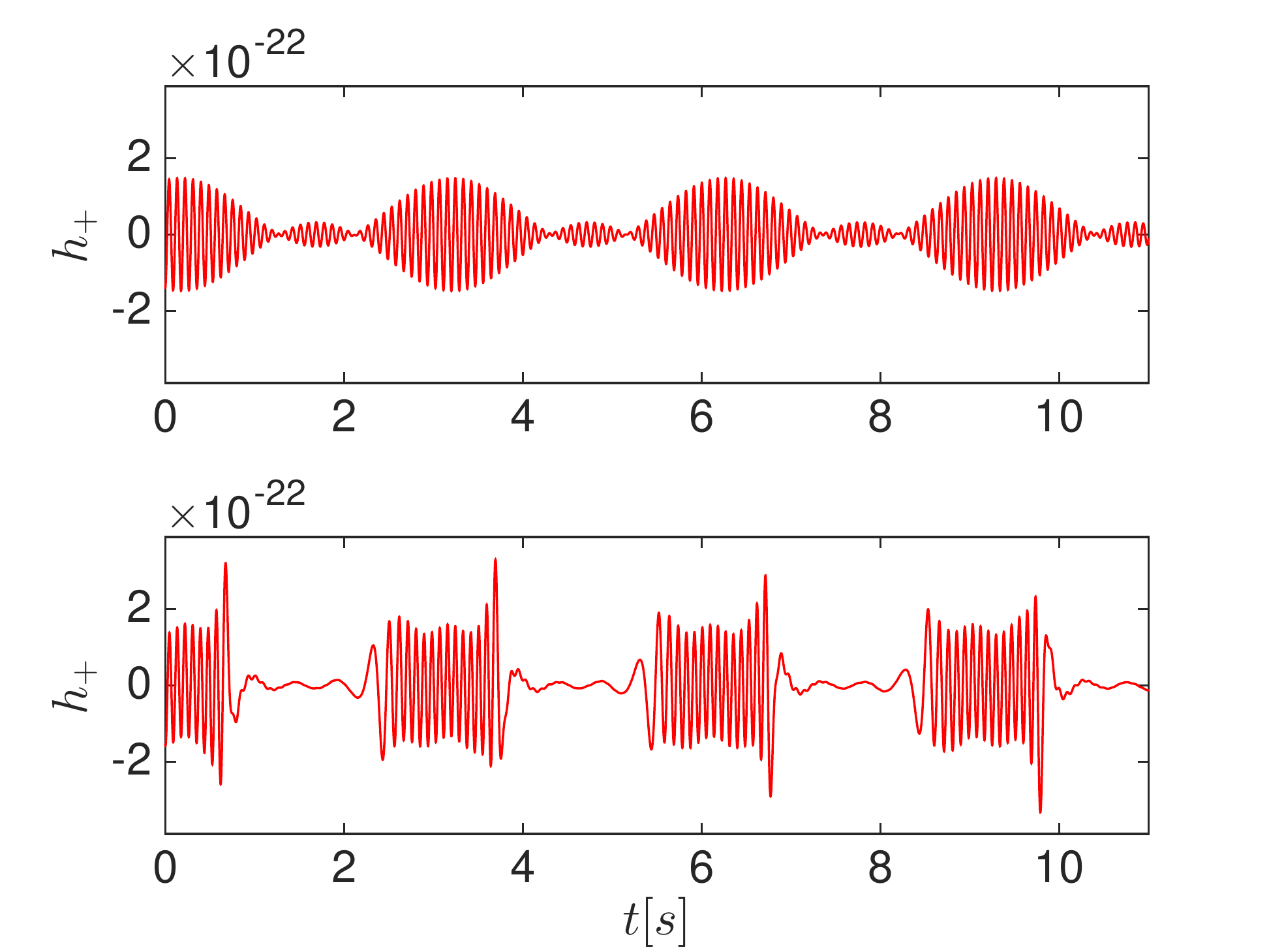}
\caption{Left: The individual harmonics or ``voices" of waveforms associated with $l=m=2$, $k=0$ (top) and $k=\tilde{k} =18$ (bottom); right: the combinations of the $k$ from 17 to 19 (top) and -10 to 30 (bottom) voices, the latter one gives the quadrupole waveform. The orbital parameters are $p=2.67$ and $e=0.8$, the dimensionless spin parameter $q=0.95$, and the mass-raio is $10^{-3}$. For plotting, the distance is normalized to 100 Mpc, and the mass of the small object is set to 2 $M_\odot$.}  \label{voice2}
\end{center}
\end{figure}

In figure \ref{voice1}, we plot two individual modes: $k=0$ and $k=\tilde{k} =10$ with $p=3.1, ~e=0.7, \nu= 10^{-3}$ and $a=0.9$ in the left panels. We can see the frequency and strength of $k=10$ mode is larger than the $k=0$ one. A mix tones of three maximum modes $k=9,~ 10,~ 11$ and $h_{22}$ waveform are shown in the right panels. One can clearly see the zoom-whirl property. Similarly, for the case of $p=2.67, ~e=0.8, \nu= 10^{-3}$ and $a=0.95$, the $\tilde{k} = 18$, then the dominant frequency is even higher than the $e=0.7$ case (see Figure. \ref{voice2}). The waveform shown in Figure. \ref{voice2} demonstrates stronger zoom-whirl behavior. All these waveforms are produced from the numerical Teukolsky-based waveforms, combined with the EOB orbits without radiation-reaction. Please see \cite{han10,han14,Han11} for the details of our numerical algorithms. 

\begin{figure}
\begin{center}
\includegraphics[height=2.0in]{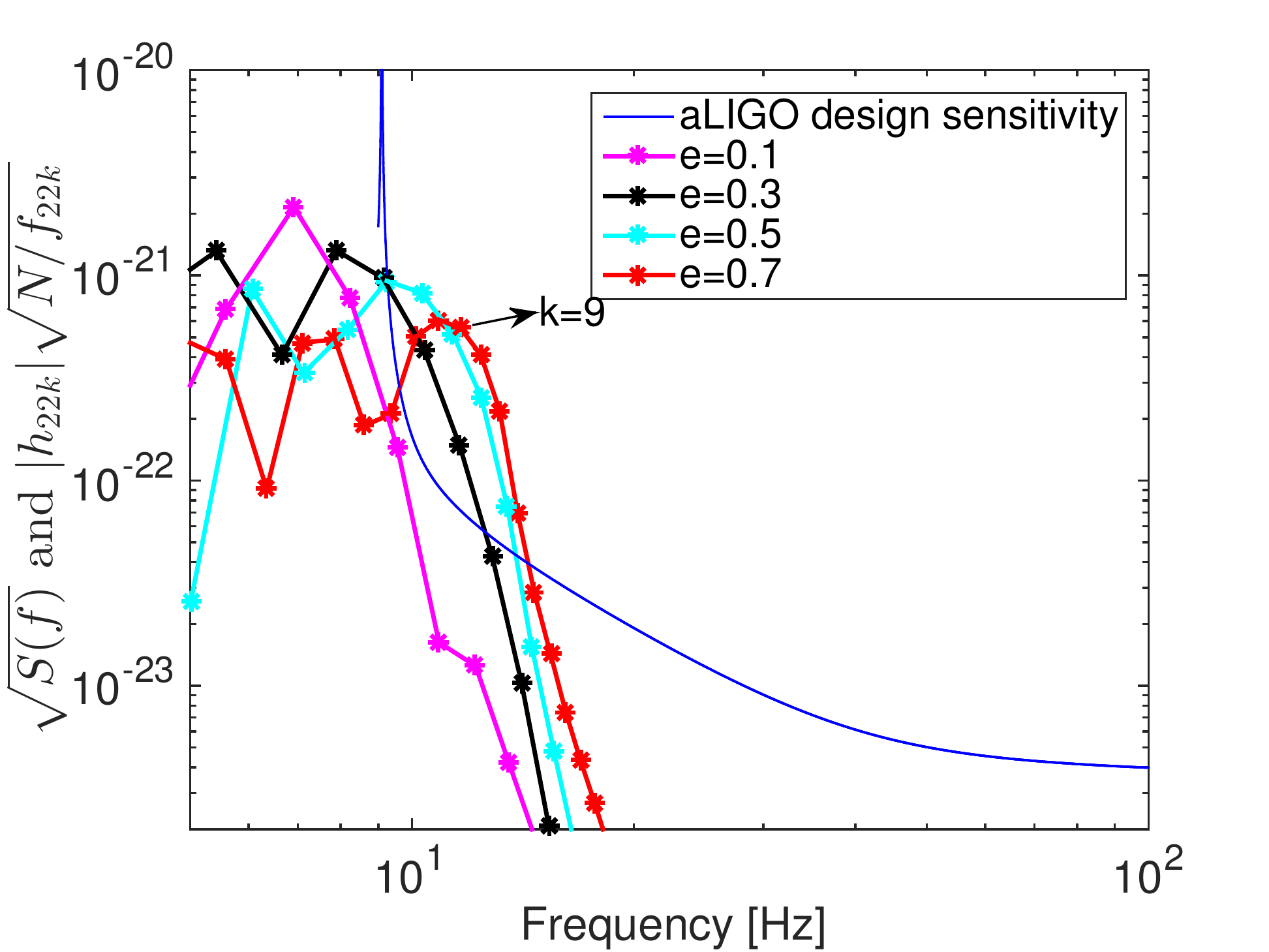}
\includegraphics[height=2.0in]{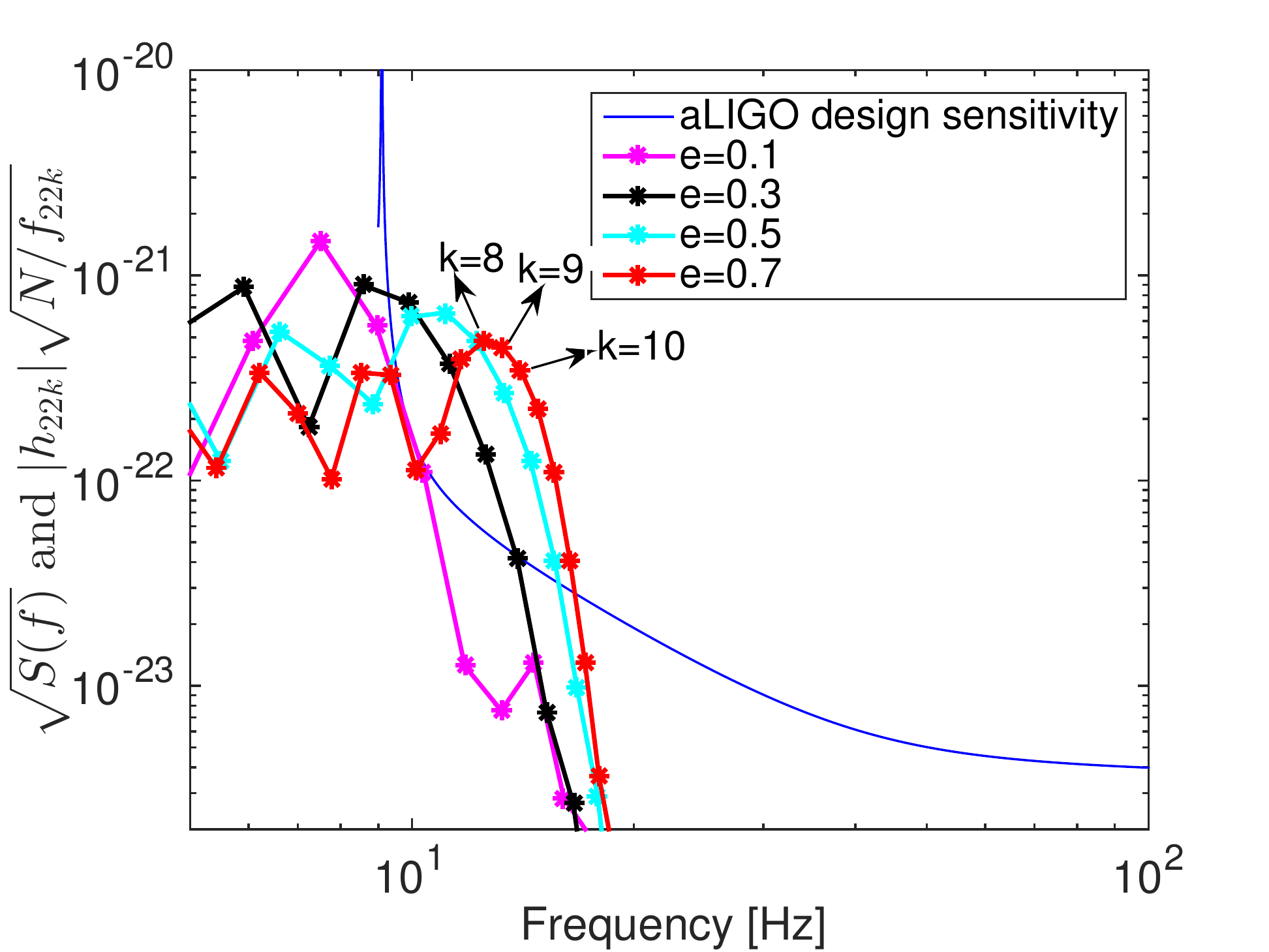}
\includegraphics[height=2.0in]{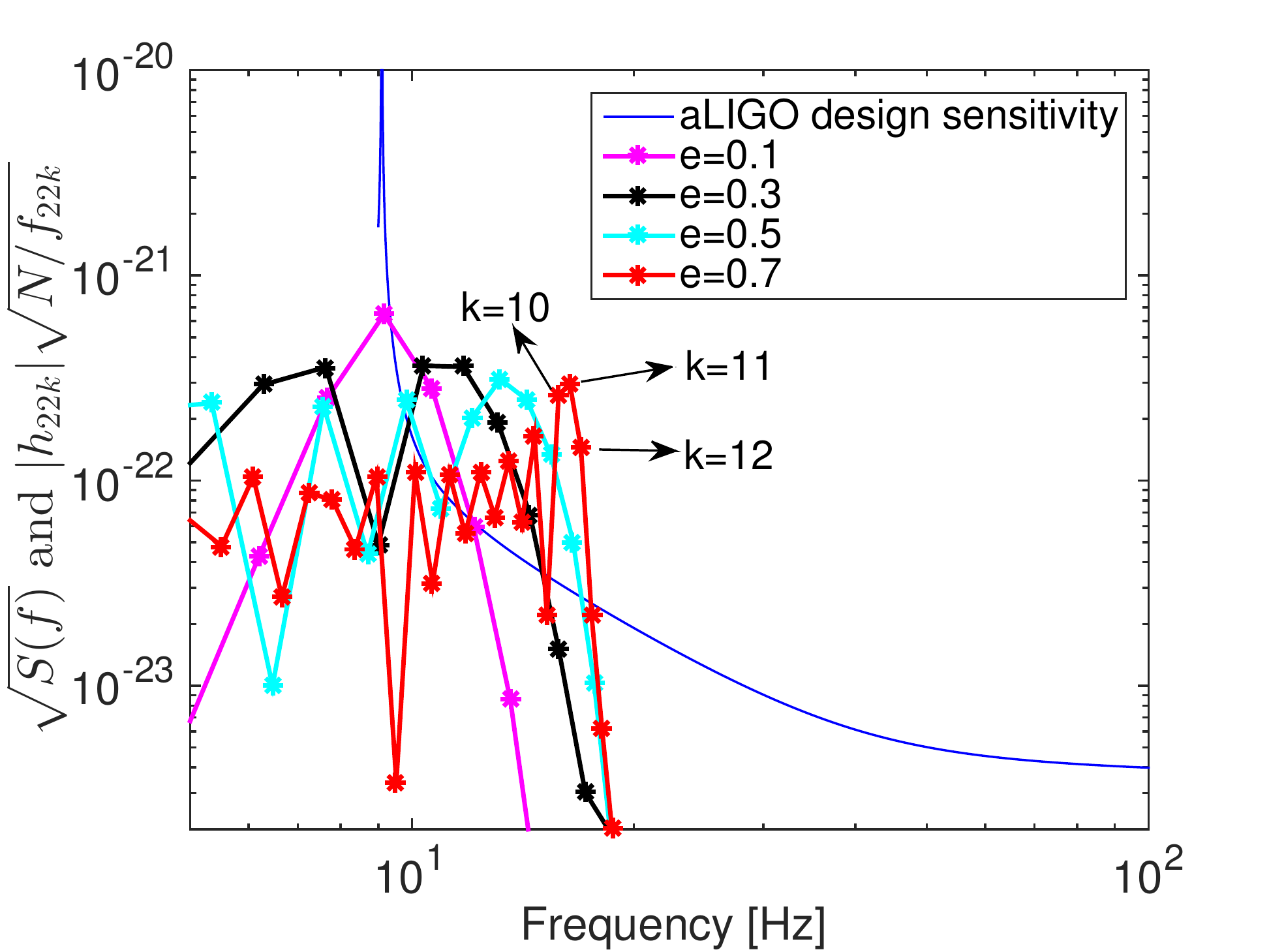}
\includegraphics[height=2.0in]{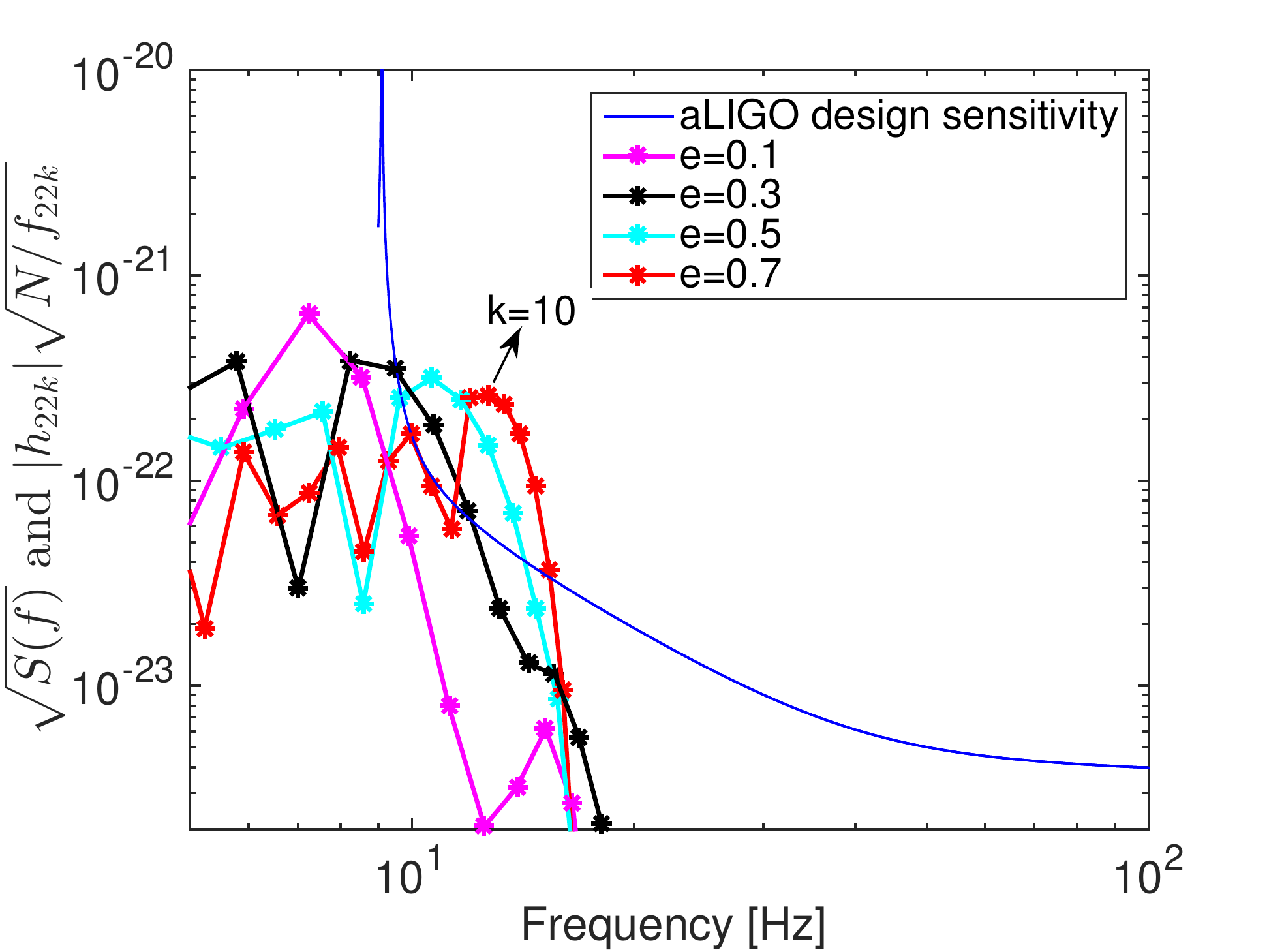}
\caption{Amplitude spectral density of the total strain noise of the default aLIGO design sensitivity, $\sqrt{S(f)}$, in units of strain per $\sqrt{\rm Hz}$, and the GW signals associated with the individual harmonics $l=m=2$ and $k$ of a group of IMRI systems plotted so that the relative amplitudes is related to the SNR of the signal. The GW sources are located at a fiducial distance of 100 Mpc, the observer is at the equatorial plane of the massive black hole.  Top-left: $q=0.95$, $p=2.6 M$, $\mu = 9 M_\odot$, $M = 1800 M_\odot$ with different orbital eccentricities, and the signal length is 17.74 s (2000 $M$); Top-right: $q=0.9$, $p=3.0 M$, $\mu = 7 M_\odot$, $M = 1400 M_\odot$ with different orbital eccentricities, and the signal length is 13.80 s (2000 $M$); bottom-left:  $q=0.95$, $p=2.55 M$, $\mu = 1.4 M_\odot$, $M = 1400 M_\odot$ with different orbital eccentricities, and the signal length is 68.96 s (10000 $M$); bottom-right:  $q=0.9$, $p=3.1 M$, $\mu = 1.4 M_\odot$, $M = 1400 M_\odot$ with different orbital eccentricities, and the signal length is 68.96 s (10000 $M$). } \label{ligo1}
\end{center}
\end{figure}

Due to this excitation mechanism, the frequencies of GWs from highly eccentric IMRIs with $\gtrsim 1,000 M_\odot$ can enter into the sensitive band of aLIGO design sensitivity (\cite{ligo}, $\gtrsim$ 10 Hz). In figure.  \ref{ligo1}, the strains of individual GW voices of several group of IMRIs and the total strain noise of the aLIGO detectors are plotted. Following \cite{moore15}, we calculate the characteristic strain of the GW source as
\begin{align}
\sqrt{S_h(f)} = |h_{22k}| \sqrt{N/f_{22k}} \,, 
\end{align}
where $N$ and $f_{22k}$ mean the number of cycles and the frequency of the $h_{22k}$ voice respectively. The total masses of the four kinds of IMRIs are set to 1400 and 1800$M_\odot$, the symmetric mass-ratios are $1 \times 10^{-3}$ and 0.005, respectively, and the distance of source from the Earth is 100 Mpc. We demonstrate the shift of GW modes from the lower frequency to the higher frequency while the eccentricity becomes larger. 

When $e \lesssim 0.1$, the dominant voice of gravitational waves is the harmonic $l=m=2, ~k=0$ mode.  Therefore, the dominant frequency of GWs is simply $(2\Omega_\phi + 0 \Omega_r)/2\pi$. It means that the frequency of the radial motion does not contribute to the dominant one of GWs. From Figure \ref{ligo1}, the pink curves in all panels show the frequency of dominant modes, and they are the lowest comparing to the others. Though the strength of dominant voices in $e=0.1$ cases are the strongest compared to the larger eccentricity orbits, due to the lowest frequency, they are well below the noise of aLIGO detectors \cite{ligo}.  The signal-noise ratios (SNR) of the $e=0.1$ orbits are very small (see Tab. \ref{snrgw} for details).  

\begin{table}[h!]
\caption{The maximum SNRs of $h_{22k}$ of the four systems in Fig. \ref{ligo1}. The Sys I to IV respect to the systems from top to bottom, left to right panels. ``$--$'' means the SNR $\ll$ 1.}
\label{snrgw}
\begin{center}
\begin{tabular}{c| c c c c c}
\hline \hline
 &$e=0.1$ & $e=0.3$ & $e=0.5$ & $e=0.7$  \\
\hline
Sys I & $--$ & $3.6$ & $6.5$ & $7.8$  \\
Sys II & $--$ & $4.6$ & $7.8$ & $9.2$  \\
Sys III & $2.7$ & $5.2$ & $6.3$ & $10.0$  \\
Sys IV & $--$ & $1.8$ & $3.5$ & $5.1$  \\
\hline \hline
\end{tabular}
\end{center}
\end{table}

\begin{figure}
\begin{center}
\includegraphics[height=2.0in]{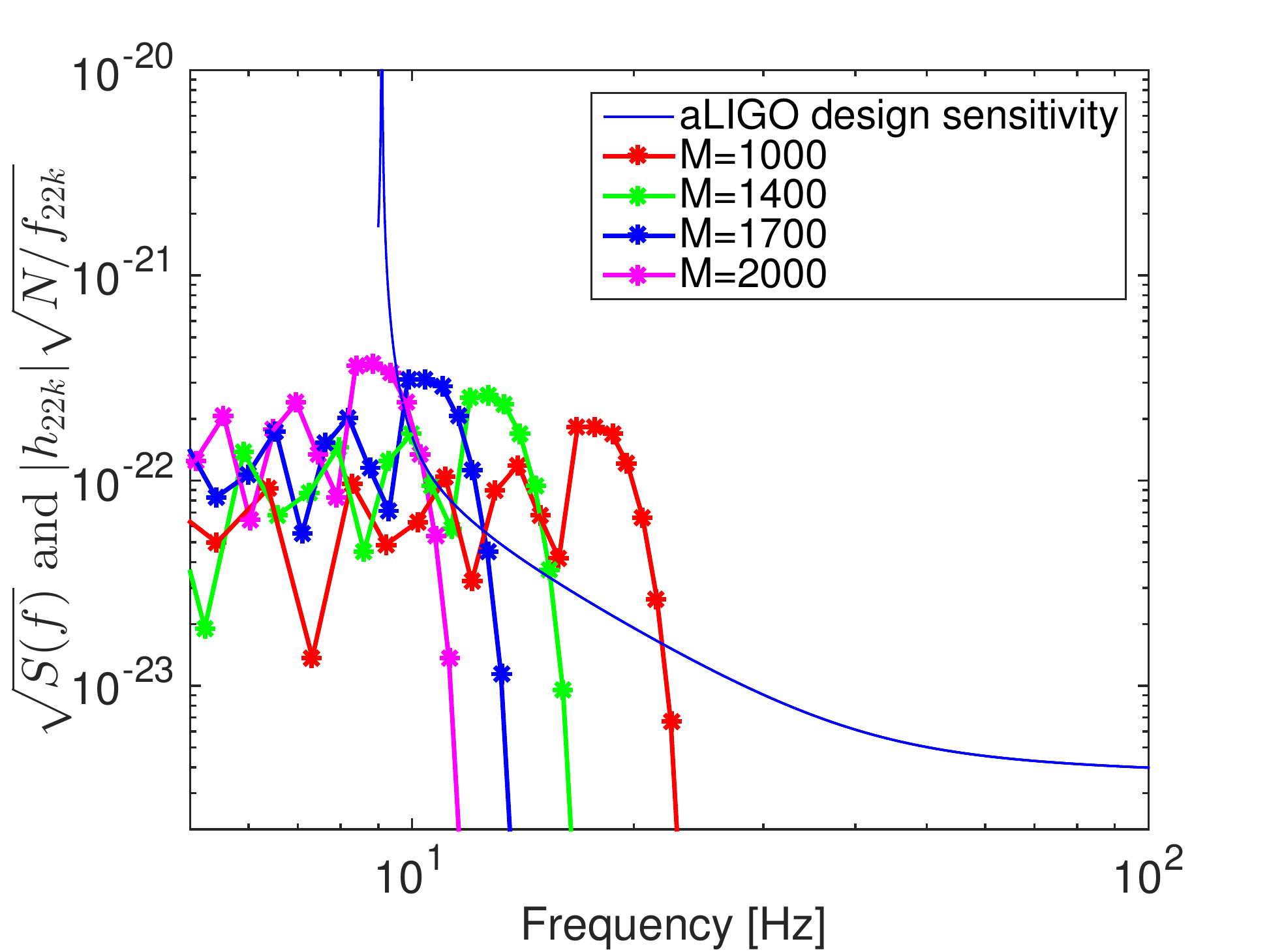}
\includegraphics[height=2.0in]{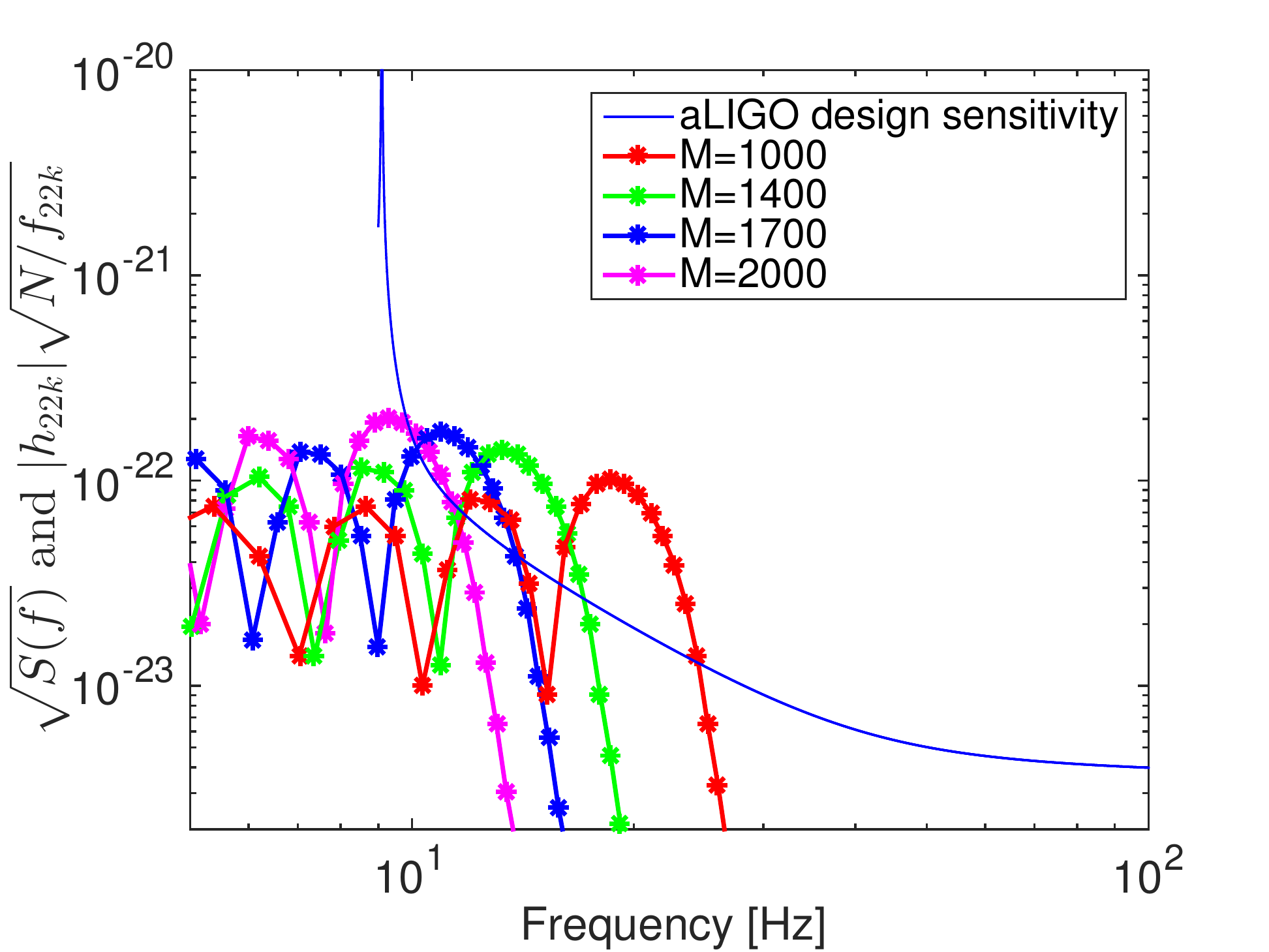}
\includegraphics[height=2.0in]{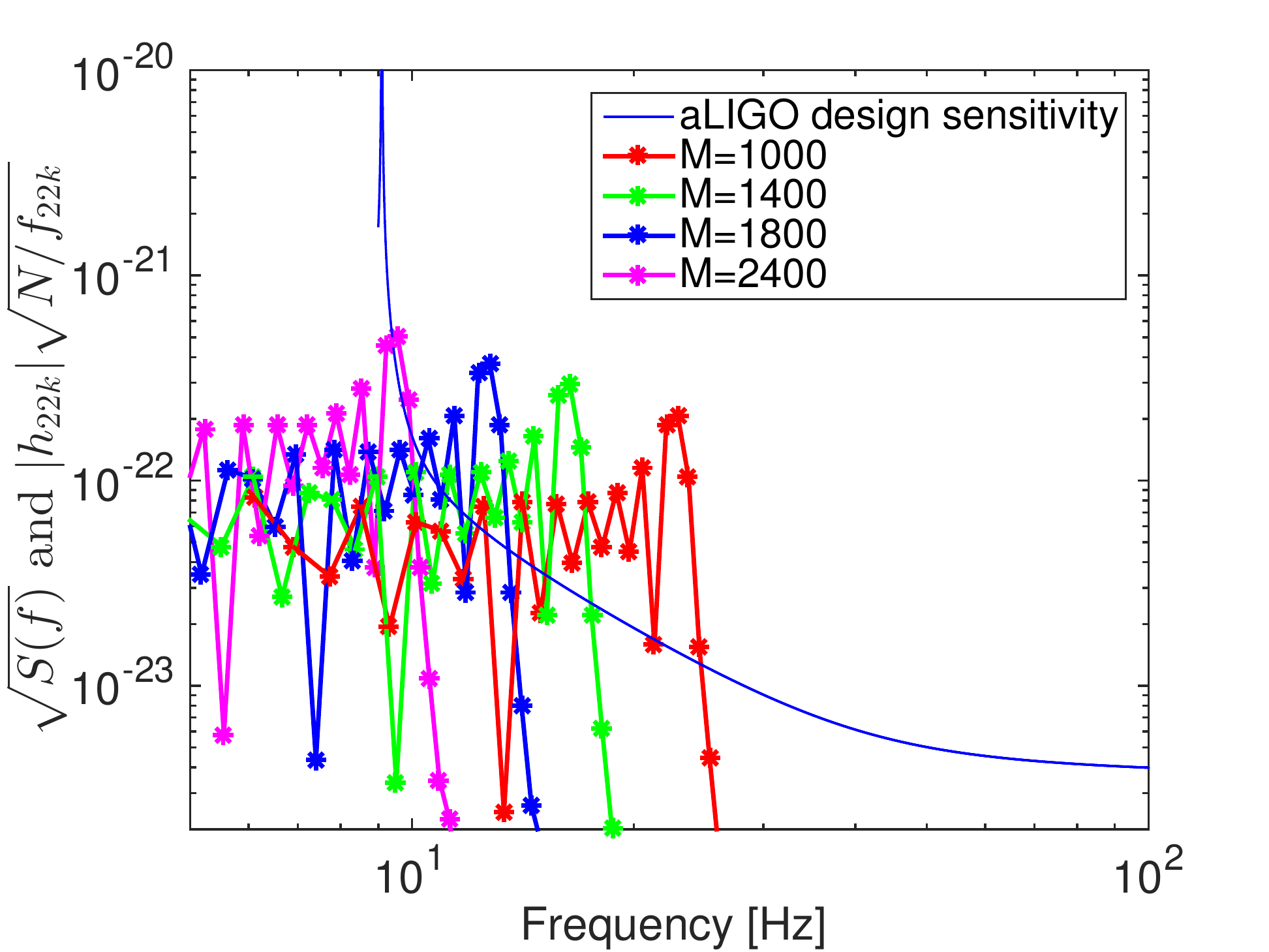}
\includegraphics[height=2.0in]{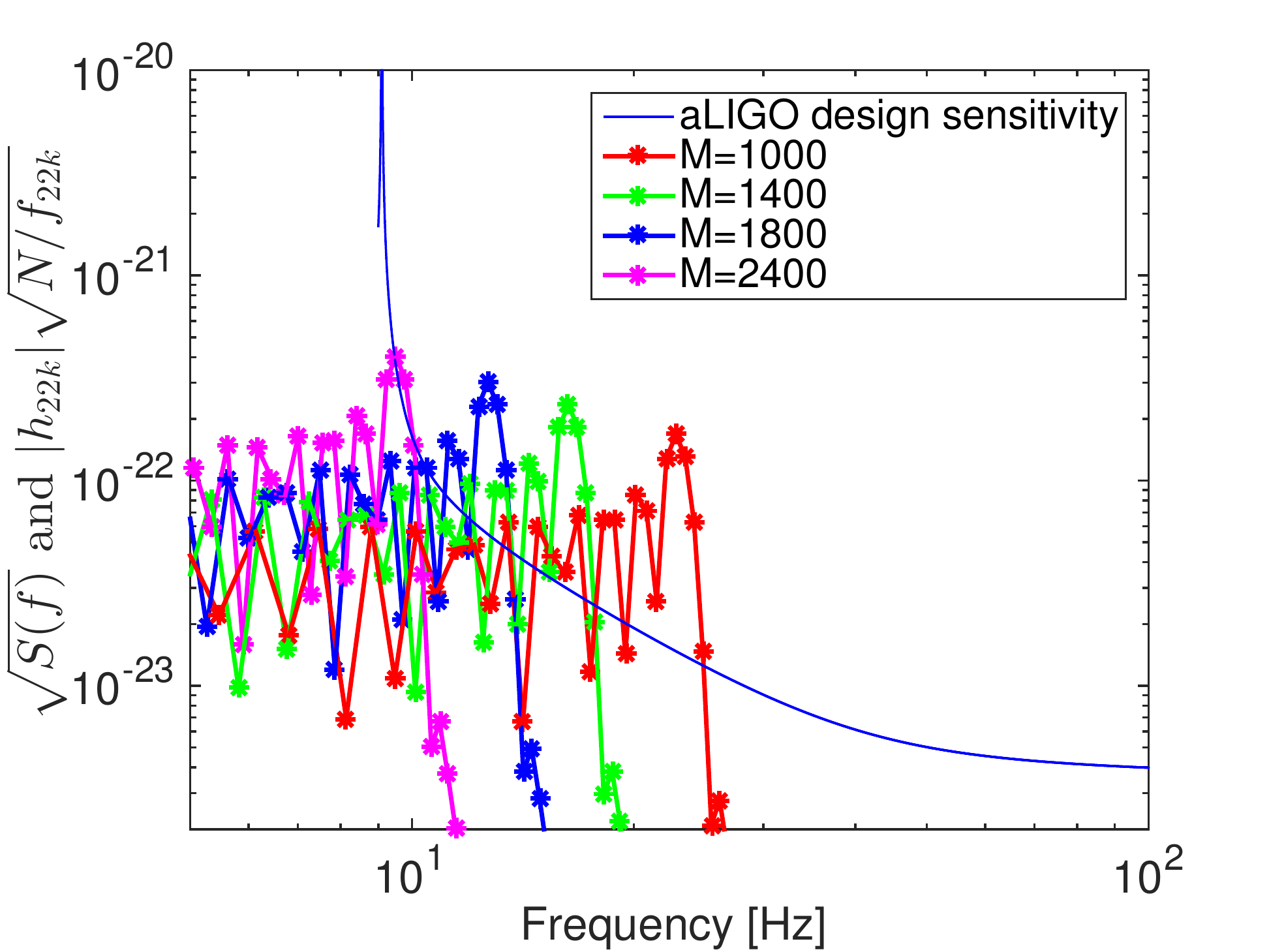}
\caption{Amplitude spectral density of the total strain noise of the default aLIGO design sensitivity, $\sqrt{S(f)}$, in units of strain per $\sqrt{\rm Hz}$, and the GW signals associated with the individual harmonics $l=m=2$ and varied $k$ of a group IMRI systems with mass-ratio $\mu/M = 10^{-3}$ plotted so that the relative amplitudes is related to the SNR of the signal. The GW sources are located at a fiducial distance of 100 Mpc, the observer is at the equatorial plane of the massive black hole.  Top-left: $q=0.9$, $e=0.7$, $p=3.1 M$, about 46 orbits; top-right:  $q=0.95$, $e=0.8$, $p=3 M$, about 40 orbits; bottom-left: $q=0.95$, $e=0.7$, $p=2.55 M$, about 40 orbits; bottom-right:  $q=0.95$, $e=0.8$, $p=2.67 M$, about 34 orbits. The unit of $M$ in the panels is the $M_\odot$.} \label{ligo2}
\end{center}
\end{figure}

When eccentricity goes larger, the strength of dominant voices becomes weaker. This is because in the case of highly eccentric orbit, more fluxes distribute to the non-dominant modes. However, as the dominant modes now locate at $k > 0$, for example, in the $e=0.7$ cases, $\tilde{k} = 10$, then the dominant frequency is $(2\Omega_\phi + \tilde{k} \Omega_r)/2\pi$. Though the $\Omega_\phi$ becomes smaller when $e$ goes larger, with the contribution of $\Omega_r$, the frequency of dominant voice still become higher while the eccentricity goes larger, and move into the sensitive band, then the SNR still increase as the eccentricities go higher. Particularly for the cases $e=0.7$,  we can see from the Tab.  \ref{snrgw}, the SNR can be as large as 10. Comparing with the $e=0.1$ cases, the SNRs are obviously larger. Except for the bottom-left panel (the largest SNR of $e= 0.7$ case is also 5.1), the systems with $e=0.7$ are all associated with considerable SNRs. We may conclude that the highly eccentric IMRIs are more valuable sources than the lowly eccentric counterparts for ground-based GW detectors like advanced LIGO if they have the same semi-latus rectums. 

Be careful, the largest SNRs we show in Tab. \ref{snrgw} do not always correspond to the highest modes. This is due to the sharply decreasing noise curve of detector with the frequency increasing after 10 Hz. 

Obviously, the GW frequency mainly depends on the mass of the IMRIs. If the total mass is too large,  the
frequency of GWs will be too low to be detected by aLIGO or AdV.  For revealing this point, we use four IMRI systems with different total masses to demonstrate the detectability of the highly eccentric binaries. For IMRIs with total mass larger than 2000 $M_\odot$, even an eccentricity of $0.8$ is still hard to excite a strong enough signal to be detected (see the figure \ref{ligo2} for details). Meanwhile, when the mass is less than 1000 $M_\odot$, even for the circular orbit, the GW frequency is high enough to enter the sensitive band of aLIGO and AdV.  So the excitation mechanism works in such a ``gray zone" (the system mass between 1000 - 2000 $M_\odot$), where the circular orbit's GWs are difficult to be observed but with the help of the large eccentricity, the excited high frequency modes can enter the sensitive band of aLIGO and AdV. 

Noting that in the figures \ref{ligo1}, \ref{ligo2}, the characteristic strain is plotted by assuming the source without inspiralling, so we get just a point  for each single mode. This is approximately correct if the mass-ratio is small and the integration time is short. 


\section{Conclusions}
In this paper, we discuss the detectability of the harmonic modes of GWs from IMRIs with large eccentricity.  Compared with small eccentricity cases, the GWs from highly eccentric IMRIs can have larger SNRs for the ground-based GW detectors like aLIGO and AdV. As demonstrated in the previous sections, the harmonic number $k$ of the dominant mode goes larger as the eccentricity gets higher.  As a result, the frequency of the dominant GW mode of high eccentricity cases can be higher than the small eccentricity cases with the same semi-latus rectums.  We call this mechanism as an excitation of high frequency GWs by eccentricity. For a group of IMRIs with appropriate masses, this mechanism can shift the GW signals into the sensitive band of aLIGO/AdV detectors, making it possible for such systems to be detected.   

More precisely, the excitation mechanism by the eccentricity can make some IMRIs with mass around 1000-2000 $M_\odot$ become detectable for aLIGO/AdV. While these IMRIs are out of the sensitive band of aLIGO/AdV if they are circular or small eccentricity orbits.  The IMRIs with mass more than 1000 $M_\odot$ and less than 2000 $M_\odot$ are in a gray zone of ground-based and space-based GW detectors aLIGO, AdV, LISA, Taiji and Tianqin. We argue here that these IMRIs still have an opportunity to be found by aLIGO/AdV if their eccentricities are as high as $0.7$. If detected, it will have a great impact on our understanding of astrophysics. For example, the significant eccentricities of IMRIs are indicative of  the direct capture scenario via two-body relaxation \cite{pau07}.

\section*{Acknowledgement} 
 This work is supported by NSFC No. 11773059 and U1431120, QYZDB-SSW-SYS016 of CAS; NSFC No. 11690023, 11622546 and 11375260. WH is also supported by Youth Innovation Promotion Association CAS. We appreciate for the useful comments from Dr. Tito.

{}   \end{document}